\documentclass[aps,pra,reprint,superscriptaddress,longbibliography]{revtex4-2}
\usepackage{orcidlink} 
\usepackage{graphicx}
\usepackage{dcolumn}
\usepackage{bm}
\usepackage{amsthm}
\usepackage{multibib}

\usepackage{braket}
\usepackage{amsmath}
\usepackage{bibentry}
\usepackage{color,soul}

\newcommand{\nocontentsline}[3]{}
\newcommand{\tocless}[3]{\bgroup\let\addcontentsline=\nocontentsline#1{#3}\egroup}

\AtBeginDocument{\let\oldcontentsline\contentsline}
\newcommand{\notoccontentsline}[4]{\oldcontentsline{}{}{}{}}
\newcommand{\droptocpage}{\addtocontents{toc}{\let\protect\contentsline\protect\notoccontentsline}}
\newcommand{\incltocpage}{\addtocontents{toc}{\let\protect\contentsline\protect\oldcontentsline}}

\begin{document}
	\title{Microscopic Study on Superexchange Dynamics of Composite Spin-1 Bosons}
	
	\author{An Luo}
	\thanks{These authors contributed equally to this work.}
	\author{Yong-Guang Zheng}
	\thanks{These authors contributed equally to this work.}
	\author{Wei-Yong Zhang}
	\author{Ming-Gen He}
	\author{Ying-Chao Shen}
	\author{Zi-Hang Zhu}
	
	\affiliation{Hefei National Research Center for Physical Sciences at the Microscale and School of Physical Sciences, University of Science and Technology of China, Hefei 230026, China}
	\affiliation{CAS Center for Excellence in Quantum Information and Quantum Physics, University of Science and Technology of China, Hefei 230026, China}

	\author{Zhen-Sheng Yuan}
	
	\affiliation{Hefei National Research Center for Physical Sciences at the Microscale and School of Physical Sciences, University of Science and Technology of China, Hefei 230026, China}
	\affiliation{CAS Center for Excellence in Quantum Information and Quantum Physics, University of Science and Technology of China, Hefei 230026, China}
	\affiliation{Hefei National Laboratory, University of Science and Technology of China, Hefei 230088, China}
	
	\author{Jian-Wei Pan}
	
	\affiliation{Hefei National Research Center for Physical Sciences at the Microscale and School of Physical Sciences, University of Science and Technology of China, Hefei 230026, China}
	\affiliation{CAS Center for Excellence in Quantum Information and Quantum Physics, University of Science and Technology of China, Hefei 230026, China}
	\affiliation{Hefei National Laboratory, University of Science and Technology of China, Hefei 230088, China}

	\date{\today}
	
	\begin{abstract}
		We report on an experimental simulation of the spin-1 Heisenberg model with composite bosons in a one-dimensional chain based on the two-component Bose-Hubbard model. 
		Exploiting our site- and spin-resolved quantum gas microscope,
		we observed faster superexchange dynamics of the spin-1 system compared to its spin-1/2 counterpart, 
		which is attributed to the enhancement effect of multi-bosons.
		We further probed the non-equilibrium spin dynamics driven by the superexchange and single-ion anisotropy terms, 
		unveiling the linear expansion of the spin-spin correlations, which is limited by the Lieb-Robinson bound.
		Based on the superexchange process, we prepared and verified the entangled qutrits pairs with these composite spin-1 bosons, potentially being applied in qutrit-based quantum information processing.
		
	\end{abstract}
	
	\maketitle
	\droptocpage
	
	\textbf{\textit{Introduction.---}}
	Ultracold atoms in optical lattices constitutes a powerful tool for investigating intricate magnetic phases and exploring spin dynamics for its high degree of isolation, controllability, and detection \cite{greiner2002quantum,bloch2008many,gross2017quantum}.
	Previous studies have shown that a two-component Mott insulator in an optical lattice can be used to realize the spin-1/2 Heisenberg model \cite{duan2003controlling,kuklov2003counterflow,trotzky2008time,Fukuhara2013b,sun2021realization} and 
	to generate atomic entanglement based on spin exchange process \cite{dai2016generation,Yang2020,zhang2023scalable}. 
	When two spin-1/2 particles co-occupy a single lattice site, they form a composite spin-1 bosons, thereby leading the system to the spin-1 regime \cite{kuklov2003counterflow,altman2003phase}.
	In this case, the single-ion anisotropy term, $u(S^z_i)^2$, \textcolor{black}{also} emerges.
	This term explicitly breaks the SU(2) symmetry of the system into the residual U(1) symmetry in the xy-plane, playing an important role in stabilizing magnetism. 
	The spin-1 Heisenberg model shows a range of intricate phenomena, such as the XY ferromagnetic phase and the Haldane phase \cite{haldane1983nonlinear,golinelli1993magnetic,altman2003phase,langari2013ground}, and it has attracted widespread interests both theoretically \cite{li2011tunable,wierschem2012magnetic,kennedy1992hidden,hamer2010spin,schachenmayer2015adiabatic} and experimentally \cite{rogado2002bani,zvyagin2007magnetic,chung2021tunable,de2022preparation,chauhan2020tunable,sompet2022realizing}.
	Furthermore, from the viewpoint of quantum information processing, this controllable spin-1 system posses greater channel capacity and enhanced robustness to noise \cite{fujiwara2003exceeding,groblacher2006experimental,lanyon2008manipulating} in comparison to the spin-1/2 system, underscoring the necessity of investigating such higher spin systems.
	
	In the simulation of spin-1 Heisenberg model in an optical lattice, a pivotal aspect lies in the dynamic interplay between the spin exchange interaction
	, driven by higher-order virtual processes, 
	and the single-ion anisotropy term, originating from the difference between the inter- and intra-component interactions \cite{kuklov2003counterflow,altman2003phase}. These factors can be controlled by tuning the parameters of the optical lattice.
	However, the study of such spin-1 system faces two significant challenges. Firstly, preparing low-entropy initial states is imperative \cite{chiu2018quantum,Yang2020}, as an undesired filling may lead to particle doping in other spin spaces, thereby disrupting the spin-1 dynamics. Secondly, achieving a site-resolved readout of the spin-1 state poses a challenge, since it demands simultaneous resolution of both particle number and spin state.

	\begin{figure}[htb]
		\includegraphics[width=0.99\linewidth]{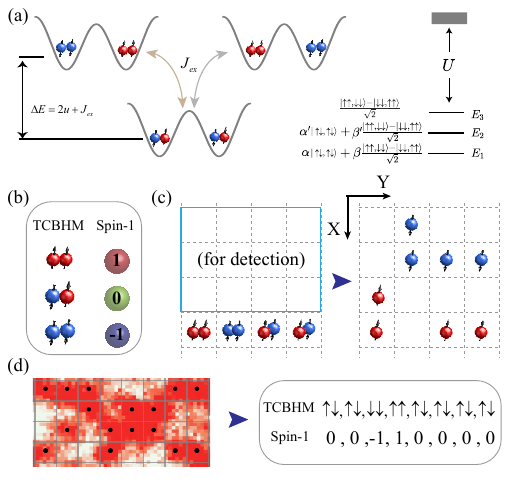}
		\caption{\label{fig:setup}Preparation and detection of the spin-1 composite bosons. 
			(a) Schematics of the superexchange process. The superexchange $J_{ex}$ is generated by the exchange of $\ket{\uparrow}$ and $\ket{\downarrow}$ via second-order process. it couples $\ket{0,0}$ to $\ket{1,-1}$ and $\ket{-1,1}$. there is a energy shift $\Delta E = 2u+J_{ex}$ of $\ket{0,0}$ and $\ket{1,-1}$ ($\ket{-1,1}$). Here, $u$ is the single-ion isotropic intensity of the spin-1 Heisenberg model. On the right are the three eigan states of the two site model, the respected energy are $E_1 = (u+J_{ex}/2)-\sqrt{(u+J_{ex}/2)^2+2J_{ex}^2}$, $E_2 = (u+J_{ex}/2)+\sqrt{(u+J_{ex}/2)^2+2J_{ex}^2}$ and $E_3 = 2u+J_{ex}$, there is a energy gap $U$ from those state to the excited state.
			(b) The mapping relationship between the two components Bose-Hubbard model (TCBHM) and the spin-1 Heisenberg model. 
			(c) The spin detection procedure. The spin-1 (1/2) chain is created along $y$-direction, and four lattice sites along $x$-direction are used for detection. During the detection process, the $\ket{\downarrow}$ atoms are moved to the top two sites while the $\ket{\downarrow}$ atoms are confined in the two bottom sites.
			(d) An exemplary fluorescence image in a single experimental realization. The black dot is the reconstructed atomic distribution, and the right panel shows the deduced status in both TCBHM (upper) and spin-1 (bottom) presentations.}
	\end{figure}

	In this Letter, we report on the realization of a spin-1 Heisenberg model using a two-component Mott insulator in optical lattices.
	This is achieved by combining parallel and local atomic spin manipulations in the spin-dependent superlattices \cite{Yang2017}. 
	We employed a site- and spin-resolved quantum gas microscope to directly monitor the real-time superexchange dynamics of both spin-1/2 and spin-1 systems in the double-well potentials. 
	In comparison with the spin-1/2 case, faster superexchange dynamics for the spin-1 situation was observed, which is attributed to the enhancement effect of multi-bosons. 
	By halting the superexchange-driven evolution of the spin-1 systems in the double well, 
	we created and verified the entanglement of two spin-1 particles. 
	Furthermore, we probed the time- and site-resolved non-equilibrium spin transport in a 6-site spin-1 chain,
	which is governed by the interplay between the superexchange and anisotropy. 
	We observed a light-cone-like propagation of the spin correlations, 
	\textcolor{black}{which is associated with the Lieb-Robinson bounds in a short-range interacting spin system }\cite{Lieb1972,Cheneau2012}.
	
	\textbf{\textit{Spin-1 Heisenberg models in optical lattices.---}}Our work is focused on exploring the spin-1 Heisenberg model on a one-dimensional (1D) chain, described by the following Hamiltonian
	\begin{equation}
		\label{eq_s1Heisenberg}
		\begin{split}
			\hat{H} = -J_{ex}\sum_{\langle i,j \rangle} \hat{\mathbf{S}}_i \cdot \hat{\mathbf{S}}_j + u\sum_{i}(\hat{S}_i^z)^2,
		\end{split}
	\end{equation}
	where, $J_{ex}$ denotes the Heisenberg exchange strength, $u$ represents the uniaxial single-ion anisotropic intensity,
	$\hat{S}_i^{\gamma}$ is the $\gamma$ ($=x,y,z$) component of spin-1 operators on site $i$, and $\langle i,j \rangle$ denotes the nearest neighbor coupling.  
	This Heisenberg model (Eq.~\ref{eq_s1Heisenberg}) can be implemented in a two-component Bose-Hubbard model (TCBHM) that describes two-component bosons in the lowest Bloch band of the lattice. 
	In the TCBHM, we label the inter-component on-site energies of spin $\ket{\uparrow}$ and spin $\ket{\downarrow}$ sites as $U_{\uparrow \downarrow}$, the tunneling amplitudes of each species as $J$, and the intra-component interactions as $U_{\uparrow \uparrow}$ and $U_{\downarrow \downarrow}$. 
	In the regime of large on-site interaction, where $U_{\uparrow \uparrow} = U_{\downarrow \downarrow} \equiv U,U_{\uparrow \downarrow} \gg J$,  the two-species Mott insulator with two atoms per site can be effectively represented by a spin-1 model in second-order perturbation theory \cite{altman2003phase}.
	\begin{figure}[tb]
		\includegraphics[width=1\linewidth]{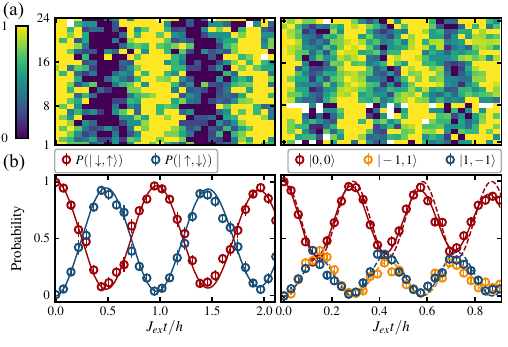}
		\caption{\label{fig:Supexchange}Superexchange processes in spin-1/2 and spin-1 system. 
			(a) The site-resolved oscillation of the initial states in spin-1/2 (left panel) and spin-1 (right panel) system in a region of $3\times 8$. 
			(b) The time-resolved averaged populations for different states during the superexchange process in the spin-1/2 (left panel) and spin-1 (right panel) system within a total of 24 double wells, the initial state is $\ket{\downarrow,\uparrow}$ for spin-1/2 and $\ket{0,0}$ for spin-1 system. Solid lines are the fitted curve with a damped sinusoidal function, dashed lines are the numeric calculation result with spin-1 Heisenberg model. The evolving time is rescaled with the fitted $J_{ex}t$ in unit of $2\pi$.
			Error bars in (b) denote the standard error of the mean (s.e.m.) and are smaller than the points if not visible. Only samples in the spin state $\{\ket{\downarrow,\uparrow},\ket{\uparrow,\downarrow}\}$ and $\{\ket{1,-1},\ket{0,0},\ket{-1,1}\}$ are adopted.}
	\end{figure}
	Using this implementation, as depicted in Fig.~\ref{fig:setup}(a), the Heisenberg exchange strength and single-ion anisotropic intensity correspond to $J_{ex} = 4(J+3T)^2/U_{\uparrow \downarrow}$ ($J_{ex} = 4(J+T)^2/U_{\uparrow \downarrow}$ for $S=1/2$) and $u=U-U_{\uparrow \downarrow}$, where $T$ represents the density-induced tunneling strength \cite{Supplement}. 
	Additionally, we can identify the operators as $\hat{S}_i^+ = \hat{a}_{i,\uparrow}^\dagger \hat{a}_{i,\downarrow}$, $\hat{S}_i^- = \hat{a}_{i,\downarrow}^\dagger \hat{a}_{i,\uparrow} $, $2\hat{S}_{i}^z = \hat{n}_{i,\uparrow} - \hat{n}_{i,\downarrow} $, where $\hat{a}_{i,\alpha}^{(\dagger)}$ is the annihilation (creation) bosonic operator at site $i$ for species $\alpha$ ($=\uparrow,\downarrow$), and $\hat{n}_{i,\alpha} = \hat{a}_{i,\alpha}^\dagger \hat{a}_{i,\alpha}$ is its corresponding number operator. 
	As shown in Fig.~\ref{fig:setup}(b), the effective states of spin-1 in the $z$ direction are proportional to $\hat{a}_{i,\uparrow}^\dagger \hat{a}_{i,\uparrow}^\dagger \ket{0}$ ($\hat{S}_{i}^z=+1$), $\hat{a}_{i,\uparrow}^\dagger \hat{a}_{i,\downarrow}^\dagger \ket{0}$ ($\hat{S}_{i}^z=0$) and $\hat{a}_{i,\downarrow}^\dagger \hat{a}_{i,\downarrow}^\dagger \ket{0}$ ($\hat{S}_{i}^z=-1$).

	\textbf{\textit{Observation of the superexchange process in spin-1/2 and spin-1 systems.---}}Our experiments begin with preparing near defect-free ultracold $^{87}\rm{Rb}$ atom arrays, as described in ref.~\cite{zhang2023scalable}. 
	We encode pseudospins with two hyperfine ground states, $\ket{\downarrow} = \ket{F=1,m_{\mathrm{F}}=-1}$ and $\ket{\uparrow} = \ket{F=2,m_{\mathrm{F}}=-2}$. 
	To study the dynamics in the spin-1/2 scheme, we follow the conventional realization by arranging the two $^{87}\mathrm{Rb}$ atoms in each isolated double-well to a N\'eel-type antiferromagnetic order $\ket{\uparrow,\downarrow}$, with a half-integer filling factor for each component (the comma separating the left and right occupations). 
	For the spin-1 scheme, we use the site-resolved addressing and superlattice techniques to deterministically create commensurate filling chains (along the $y$-direction) with an integer occupation per site for each pseudospin. 
	After the time evolution, we employ a Stern-Gerlach-type process to detect the final spin status of the atoms in each site with three additional auxiliary lattice sites along the $x$-direction, as illustrated in Fig.~\ref{fig:setup}(c). 
	During this process, the $\ket{\uparrow}$ atoms are confined in the two bottom sites while the $\ket{\downarrow}$ atoms are moved to the two top sites. 
	An exemplary fluorescence image of the detected atomic distribution in a single experimental realization is shown in Fig.~\ref{fig:setup}(d), and the corresponding spin status in both the TCBHM representation and spin-1 representation are listed in the right panel of the Figure.

	We plot the measured site-resolved (averaged) superexchange dynamics of both spin-1/2 system (left panel) and spin-1 system (right panel) over 24 double wells in Fig.~\ref{fig:Supexchange}(a) (Fig.~\ref{fig:Supexchange}(b)).
	For the spin-1/2 system, the system oscillates between states $\ket{\downarrow,\uparrow}$ and $\ket{\uparrow,\downarrow}$, with a fitted superexchange coupling strength $J_{ex}/h = 14.6\pm0.2\;\rm{Hz}$. 
	The period varies slightly across the 24 double wells due to spatial inhomogeneity and a residual spin-dependent chemical potential in the experiment. 
	In the case of spin-1 system, an oscillation between states $\ket{0,0}$ and $(\ket{1,-1} +\ket{-1,1})/\sqrt2$ is observed with $J_{ex}/h = 30.3\pm0.5\;\rm{Hz}$ and $u/h = 14.7\pm0.6\;\rm{Hz}$, agree with the numeric calculation of the spin-1 model. 
	The experiment results indicate that the spin-1 dynamics exhibit a 3.4-fold increase in oscillating frequency compared to the spin-1/2 system. 
	This increase is attributed to the enhancement effect of multi-bosons (factor of 3) and the single-ion anisotropic term $u$. 
	
	\begin{figure}
		\includegraphics[width=0.99\linewidth]{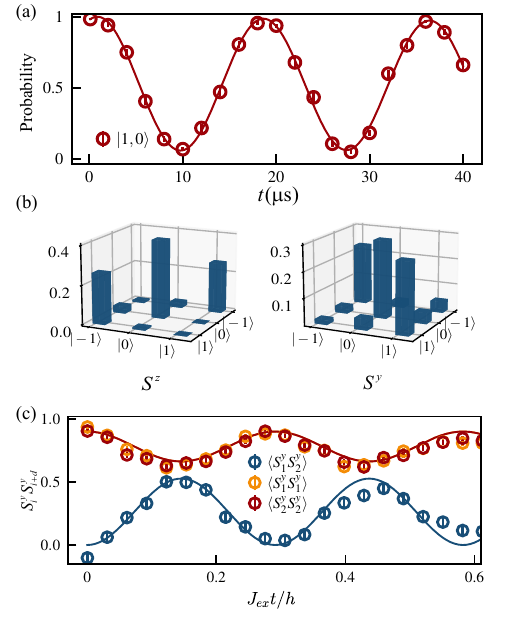} 
		\caption{\label{fig:Sy}Off-diagonal correlations and entanglemnent of qutrits. 
			(a) Microwave pulse driving oscillation of the spin-1 system. The initial state is $\ket{S,mS}=\ket{1,0}$. 
			(b) The population of different basis in $S^z$(left panel) and $S^y$(right panel) detections.
			(c) The superexchange induced oscillation of the off-diagonal correlations $\braket{\hat{S}_i^y \hat{S}_j^y}$ during the process. Solid line denote the numeric calculation of spin-1 Heisenberg model. Error bars denote the s.e.m. and are smaller than the points if not visible. We post-selected samples where each lattice site is occupied by two particles.}
	\end{figure}
	
	\textbf{\textit{Entanglement of two qutrits.}}---While major research in quantum information science has traditionally focused on qubits, the qudits (d-level quantum objects with $d>2$) is attracted a lot of interest recently \cite{jaksch2000fast,lanyon2008manipulating,fedorov2012implementation,campbell2014enhanced,Senko2015,ringbauer2022universal,goss2022high,fu2022experimental,González-Cuadra2022,cervera2022experimental}, which posses greater information storage and increased robustness to noise \cite{fujiwara2003exceeding,groblacher2006experimental,lanyon2008manipulating,kiktenko2023realization}.
	In the framework of qubits, superexchange dynamics have been proved both theoretically and experimentally to be an efficient protocol for generating scalable multipartite entanglement \cite{dai2016generation,Yang2020,zhang2023scalable}, which is a crucial resource for qubit-based quantum computing. 
	Our target composite $S=1$ bosons are well-suited to serve as qutrits (three-level quantum objects), holding the potential for qudit-based quantum computing. 
	In the following, we demonstrate the generation and verification of two qutrits entanglement through the superexchange process, which serves as a building block for scalable multi-qutrits entanglement.
	
	\begin{figure*}[t]
		\includegraphics[width=0.99\linewidth]{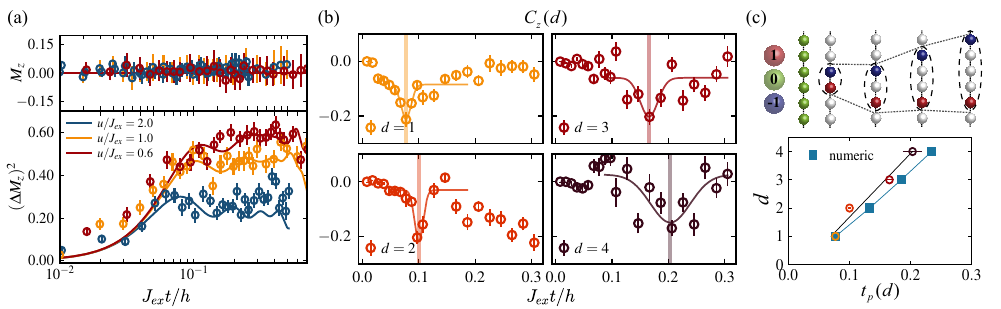}
		\caption{\label{fig:Quench}The non-adiabatic spin dynamics in spin-1 chain. 
			(a) The time evolution of the averaged local magnetization $M_z$ and its fluctuation $(\Delta M_z)^2$. The parameters are set to $u/J_{ex}=2.0,1.0,0.6$, and solid lines denote the numeric results of spin-1 Heisenberg model.
			(b) The time evolution of the spin-spin correlation $C_z(d)$ with $u/J_{ex}=1.0$ for $d=1,2,3,4$. Solid lines denote the fitted curve with a Gaussian function. The vertical shaded lines indicate the fitted times $t_p(d)$ when $C_z(d)$ reach the first peak value.
			(c) A schematic demonstration of the spreading of spin-spin correlations (upper panel). The extracted times $t_p(d)$ with varied $d$ are shown in open circles. The black line is the linear fit of the data, and the square dots are the numeric result using TEBD method. Error bars in (a,b) denote the s.e.m. and in (c) denote the fitting error. They are smaller than the circles if not visible. We post-selected samples where each lattice
			site is occupied by two particle}
	\end{figure*}
	
	In the experiment, we entangle two qutrits by halting the evolution time at the midpoint of a period during the superexchange process. To verify the entanglement, we utilize non-separability criteria characterized by the observable $W = (-1)^{\braket{\hat{S}_1^z + \hat{S}_2^z}} + 2\lvert \braket{\hat{S}^y_1 \otimes \hat{S}^y_2}\rvert$ \cite{Supplement}: two qutrits are entangled if $W > 1$. 
	To perform the $S^z$ measurement, we directly use the aforementioned Stern-Gerlach-type detection. The measured result under the $S^z$ basis is shown in the left panel of Fig~\ref{fig:Sy}(c).
	As for the off-diagonal measurement, we add an extra single qutrit rotating operator $U(\theta)$ by applying a microwave (MW) pulse: $U(\theta)\hat{S}^z U^\dagger(\theta)  = \hat{S}^z \cos\theta + \hat{S}^y \sin\theta$. 
	Fig~\ref{fig:Sy}(a) shows the MW pulse induced oscillation of the qutrit system. 
	Therefore, the $S^y$ measurement is realized by setting the duration of MW pulse to $t = 9.6\;\rm{\mu s}$. The measured result under the $S^y$ basis is shown in the right panel of Fig~\ref{fig:Sy}(c). Combining the two measurement results,
	we extracted the observable value, that is $W = 1.8 \pm 0.2$, which clearly exceeds the threshold value of 1. This result confirms the entanglement of the two qutrits.
	Moreover, we observed the time-resolved averaged off-diagonal correlations $\braket{\hat{S}^y_i \hat{S}^y_j}$ during the whole superexchange process as illustrated in Fig~\ref{fig:Sy}(b), which are in consistent with theoretical calculation based on the spin-1 model. 
	
	\textbf{\textit{Microscopic study of the non-equilibrium dynamics in spin-1 chains}}---
	In a spin-1 Heisenberg chain, the quench dynamics is governed by the interplay between superexchange and on-site anisotropy, as expressed in Eq.~\ref{eq_s1Heisenberg}.
	The first term drives the system to exchange spin between adjacent sites, while the second term acts as a stabilizing force, lowering the energy in the $\ket{0}$ state for each site.
	The competition between these terms leads to the emergence of diverse magnetic phases \cite{altman2003phase}.
	To conduct a microscopic exploration of this competing phenomenon, we prepare a 6-site spin chain initialized as $\ket{0,...,0}$, and track the out-of-equilibrium dynamics with various $u/J_{ex}$.
	Their differences are well characterized by the local magnetization $M_z = (1/L)\sum_i \braket{\hat{S}^z_i}$ and corresponding fluctuation, $(\Delta M_z)^2=(1/L)\sum_i (\braket{(\hat{S}_i^z)^2}-\braket{\hat{S}_i^z}^2)$.
	We plot measured results of $M_z$ and $(\Delta M_z)^2$ in Fig~\ref{fig:Quench}(a), and find that $M_z$ remain consistently at zero throughout the entire evolution owing to $[\hat{S}^z_i, \hat{H}]=0$, while $(\Delta M_z)^2$ grow to a finite value as a result of the spin transport. 
	Moreover, the finite value of $(\Delta M_z)^2$ is primarily influenced by $u/J_{ex}$. Our data show a clear suppression tendency of $(\Delta M_z)^2$ with an increasing $u/J_{ex}$, consistent with theoretical expectations based on the spin-1 Heisenberg model.
	
	We further investigate the spin transport behavior at a moderate ratio, $u/J_{ex}=1$, through a two-point correlation function $C^z(d) =1/(L-d)\sum_{i=1}^{L-d}(\braket{\hat{S}^z_i\hat{S}^z_{i+d}} - \braket{\hat{S}^z_i}\braket{\hat{S}^z_{i+d}})$, where $d$ is the distance between sites. 
	Fig~\ref{fig:Quench}(b) show the time evolution results of $C^z(d)$ with $d=1,2,3,4$. 
	We observe that $C^z(d)$ quickly evolves to a negative peak value, signifying the creation of $\ket{1,-1}$ pair correlations at a distance of $d$.
	A schematic understanding of this behavior is shown in the upper panel of Fig~\ref{fig:Quench}(c): The superexchange of $\ket{0,0}$ induces the formation of $\ket{1,-1}$ pairs, resulting in a negative value of $C^z(d)$ with small $d$. Subsequent superexchange process involving $\ket{1,0}$ ($\ket{-1,0}$) leads to spreading of the $\ket{1,-1}$ pairs correlations, thereby establishing correlations at larger distances $d$. 
	We extract the corresponding time $t_p(d)$ when $C_z(d)$ reaches its first negative peak, and plot them in Fig~\ref{fig:Quench}(c). 
	\textcolor{black}{A light-cone-like propagation of the spin correlations is observed, and the corresponding velocity is $v_c = 21.8 \pm 2.7$ $a_{latt}J_{ex}/h$. This is expected in a short-range interacting spin system where the velocity is bounded by the Lieb-Robinson bound \cite{Lieb1972,Cheneau2012}.}
	
	\textbf{\textit{Conclusion and outlook.}}--
	In conclusion, we experimentally study the superexchange dynamics of composite $S=1$ bosons in optical lattices. Employing state-of-the-art spin- and site-resolved manipulation and detection, we observed a Bose-enhanced superexchange process in a double well compared to the spin-1/2 case. By halting the evolution time of the superexchange process, we generated and verified two qutrits entanglement. Furthermore, we explored the non-equilibrium spin dynamics of a spin-1 Heisenberg chain, which is dominated by the interplay between superexchange and single-ion anisotropy. By extracting the spatial spin-spin correlation $C_z(d)$, we unveil a light-cone-like propagation of the spin-spin correlations.

	Our platform shows abilities for the investigation of spin-1 system. In the future, the symmetry-protected topological phases such as Haldane phase \cite{haldane1983nonlinear,haldane2017nobel,sompet2022realizing} can be studied in our system. Furthermore, by adopting the site-resolved spin flip technique, we can create spin-1 magnon and study its properties, such as the propagation of magnon or the interaction between magnons \cite{chauhan2020tunable,Sharma2021} beyond the spin-1/2 case \cite{Fukuhara2013b}. 
	
	\smallskip
	\textbf{\textit{Acknowledgements}}.---This work was supported by the NNSFC grant 12125409, the Innovation Program for Quantum Science and Technology 2021ZD0302000, and the Anhui Initiative in Quantum Information Technologies. YGZ acknowledged the support by the China Postdoctoral Science Foundation (2023TQ0102) and the CPS-Huawei MindSpore Fellowship.
	
	\bibliography{main_arXiv}    

	\onecolumngrid
	\vspace*{0.5cm}
	\newpage
	\begin{center}
		\textbf{METHODS AND SUPPLEMENTARY MATERIALS}
	\end{center}
	\vspace*{0.5cm}
	
	\twocolumngrid
	\incltocpage
	\tableofcontents
	\appendix
	\setcounter{secnumdepth}{2}
	
	\twocolumngrid
	\setcounter{equation}{0}
	\setcounter{figure}{0}
	\makeatletter
	\makeatother
	\renewcommand{\theequation}{S\arabic{equation}}
	\renewcommand{\thefigure}{S\arabic{figure}}
	\renewcommand{\thetable}{S\arabic{table}}

	\section{Extended Bose-Hubbard model and modified superexchange intensity}
	\label{appendix:c}
	Two component interacting Bose gas in optical lattice can be described by the following Hamiltonian\cite{luhmann2012multi,trotzky2008time}
		\begin{equation}
			\begin{aligned}
				\hat{H}_{\rm BHM}  = &-\sum_{i\neq j \atop \sigma}J_{ij,\sigma}\hat{a}_{i,\sigma}^\dagger \hat{a}_{j,\sigma} + \sum_{i,\sigma} \mu_{i,\sigma} \hat{n}_{i,\sigma}& \\
				\ & + \frac{1}{2}\sum_{i,j,k,l \atop \sigma_i,\sigma_j,\sigma_k ,\sigma_l}U_{i,j,k,l} \hat{a}_{i,\sigma_i}^\dagger \hat{a}_{j,\sigma_j}^\dagger \hat{a}_{k,\sigma_k} \hat{a}_{l,\sigma_l}.
			\end{aligned}
		\end{equation}
		Here, $J,\mu,U$ are the Hubbard parameters, $i,j$ are the site index, and $\sigma$ denotes the internal state index. The density induced tunneling are generated by the third term with $i=j=k$ and $l=i+1$. This term contributed as: 	
		\begin{equation}
			\hat{H}_2 =  -T\sum_{<i,j>}((N_i +N_j-1)(\hat{a}_i^\dagger \hat{a}_j +\hat{b}_i^\dagger \hat{b}_j)),
		\end{equation}
		Where $T = -U_{ia,ia,ia,ja}$, and $N_i$ ($N_j$) is the total number of site $i$ $(j)$. In this case, the final Two Components Bose-Hubbard model (TCBHM) is as follow:
		\begin{equation}
			\begin{aligned}
				\hat{H}_{\rm BHM}  = &-\sum_{<i,j> \atop \sigma}(J_{ij,\sigma}+(N_i+N_j-1))\hat{a}_{i,\sigma}^\dagger \hat{a}_{j,\sigma} \\ 
				\ & + \frac{U}{2}\sum_{i,\sigma}(\hat{n}_{i,\sigma}(\hat{n}_{i,\sigma}-1)) +
				U_{\uparrow \downarrow}\sum_i(\hat{n}_{i,\uparrow}\hat{n}_{i,\downarrow})\\
				\ & + \sum_{i,\sigma} \mu_{i,\sigma} \hat{n}_{i,\sigma}.
			\end{aligned}
		\end{equation}
		In the deep Mott region, where $U\gg J$, this Hamiltonian is mapped to the spin Heisenberg model as mentioned in the main paper. Since the total charge of each site is fixed, the superexchange term is modified as $J_{ex} = 4(J+3T)^2/U_{\uparrow \downarrow}$ for spin-1 ($J_{ex} = 4(J+T)^2/U_{\uparrow \downarrow}$ for $S=1/2$).

		\section{two-site model and generation of spin-1 GHZ state} \label{appendix:a}
		In the double well lattice structure, the system is described by the two-site Heisenberg model
		\begin{equation}
			\begin{aligned}
				\hat{H} = \left[-J_{ex}(\hat{S}_L^x\hat{S}_R^x + \hat{S}_L^y\hat{S}_R^y + \hat{S}_L^z\hat{S}_R^z)  \right] + u((\hat{S}^z_L)^2 + (\hat{S}^z_R)^2).
			\end{aligned}
		\end{equation}
		The total $\hat{S}^z=\hat{S}^z_L +\hat{S}^z_R$ is conserved during the whole dynamics since $[\hat{H},\hat{S}^z]=0$. So we only consider the subspace consist of $\{\ket{1,-1},\ket{0,0},\ket{-1,1}\}$. The Hamiltonian acting on such basis as:
		\begin{equation}
			\begin{aligned}
				\ & \hat{H}\ket{1,-1} = -J_{ex}\ket{0,0} +(2u+J_{ex})\ket{1,-1}  \\
				\ & \hat{H}\ket{0,0} = -J_{ex}(\ket{1,-1} +\ket{-1,1})\\
				\ & \hat{H}\ket{-1,1} = -J_{ex}\ket{0,0} +(2u+J_{ex})\ket{-1,1}
			\end{aligned}
		\end{equation}
		When the initial state is $\ket{0,0}$, since both the Hamiltonian and the initial state have left-right exchange symmetry, the system would oscillate between the $\ket{0,0}$ and $\frac{1}{\sqrt2}(\ket{-1,1}+\ket{1,-1})$ states. Under those two states, the Hamiltonian can be wrote as 
		\[H=\begin{pmatrix}
			0 & \sqrt2 J_{ex} \\
			\sqrt2 J_{ex} & 2u+J_{ex}
		\end{pmatrix} = \begin{pmatrix}
			-\frac{2u+J_{ex}}{2} & \sqrt2 J_{ex} \\
			\sqrt2 J_{ex} & \frac{2u+J_{ex}}{2}
		\end{pmatrix} + \frac{2u+J_{ex}}{2}\]
		The oscillating frequency and amplitude of $\ket{0,0}$ are
		\begin{equation}
			\begin{aligned}
				\ & w = 2\sqrt{2J_{ex}^2+ (\frac{2u+J_{ex}}{2})^2} \\
				\ & A = \frac{2 J_{ex}^2}{2(2J_{ex}^2+ (\frac{2u+J_{ex}}{2})^2)}
			\end{aligned}
		\end{equation}
		Under the condition that $u=J_{ex}/2$, if we halt the evolution time at $t=T/2$, the target state is $(\ket{0,0}+\ket{1,-1} + \ket{-1,1})/\sqrt{3}$, which is a highly entangle GHZ state.
		
		\section{Criterion for verifying qudit entanglement} \label{appendix:b}
		
		\textbf{Theorem.} Assume the dimension of a qudit is $D = 2S$. If an $N$-qudit quantum state $\rho$ is bi-separable for partition $M|\bar{M}$ and $i \in M, j \in \bar{M}$, the following inequality holds,
		\begin{align}
			|\left<\hat{K}_{\mathrm{total}}\right>|+\frac{1}{S^2}|\left<\hat{S}^{x}_{i}\otimes \hat{S}^{x}_j\right>|+\frac{1}{S^2}|\left<\hat{S}^{y}_{i}\otimes \hat{S}^{y}_j\right>| \le 1,
			\label{criterion}
		\end{align}
		where $\hat{K}_A = \bigotimes_{l\in A}{(-1)^{S+\hat{S}^z_l}}$ and $\hat{S}^{x}, \hat{S}^{y}, \hat{S}^{z}$ are Spin-$S$ operators.
		\begin{proof}
			The proof of this theorem is as follows.
			We first consider the situation of a pure state. If $\rho$ can be bi-separete in $M|\bar{M}$, that is $|\phi\rangle = |\phi_M\rangle|\phi_{\bar{M}} \rangle$, with $i \in M ,j\in\overline{M}$, then
			\begin{widetext}
				\begin{align*}
					& \lvert \braket{\hat{K}_{total}} \rvert + \frac{1}{S^2}\lvert \braket{\hat{S}_i^x \otimes \hat{S}_j^x} \rvert + \frac{1}{S^2}\lvert \braket{\hat{S}_i^y\otimes \hat{S}_j^y} \rvert 
					= \lvert \braket{\hat{K}_M} \rvert \lvert \braket{\hat{K}_{\bar{M}}} \rvert + 
					\frac{1}{S^2} \lvert \braket{\hat{S}_i^x} \rvert \lvert \braket{\hat{S}_j^x}\rvert+ 
					\frac{1}{S^2} \lvert \braket{\hat{S}_i^y} \rvert \lvert \braket{\hat{S}_j^y}\rvert \\
					\leq &  \sqrt{\lvert \braket{\hat{K}_M}\rvert^2 + \frac{1}{S^2} \lvert\braket{\hat{S}_i^x}\rvert^2 + \frac{1}{S^2} \lvert\braket{\hat{S}_i^y}\rvert^2} \times 
					\sqrt{\lvert \braket{\hat{K}_{\bar{M}}}\rvert^2 + \frac{1}{S^2} \lvert\braket{\hat{S}_j^x}\rvert^2 + \frac{1}{S^2} \lvert\braket{\hat{S}_j^y}\rvert^2}
				\end{align*}
			\end{widetext}
			where the inequality is based on Cauchy-Schwarz inequality.
			Next, we will prove the following inequality,
			\begin{align}
				{\left<\hat{K}_{M}\right>}^2+\frac{1}{S^2}{\left<\hat{S}_i^x\right>}^2+\frac{1}{S^2}{\left<\hat{S}_i^y\right>}^2 \le 1.
				\label{eqn:key}
			\end{align}
			Let denote $k=\left<\hat{K}_M\right>, x=\left<\hat{S}_i^x\right>, y=\left<\hat{S}_i^y\right>$. It is evident that inequality (\ref{eqn:key}) holds true if $x^2+y^2=0$. In the case where $x^2+y^2\neq 0$, we introduce the following operator:
			\begin{align*}
				\hat{O}_i = k \hat{K}_M + \frac{x}{S^2}\hat{S}_i^x + \frac{y}{S^2}\hat{S}_i^y = k \hat{K}_M + \frac{\sqrt{x^2+y^2}}{S^2}\hat{S}_i^{(\theta)},
			\end{align*}
			where $\hat{S}_i^{(\theta)} = \cos{\theta}\cdot \hat{S}_i^x + \sin{\theta}\cdot \hat{S}_i^y$, and
			$$
			\cos{\theta} = \frac{x}{\sqrt{x^2+y^2}},~~\sin{\theta} = \frac{y}{\sqrt{x^2+y^2}}.
			$$
			Consequently, the expression for $\hat{O}_i^2$ becomes:
			\begin{widetext}
				\begin{align*}
					\hat{O}_i^2 = k^2 \hat{I}_M + \frac{x^2+y^2}{S^4}\left[\hat{S}_i^{(\theta)}\right]^2 + \frac{k\sqrt{x^2+y^2}}{S^2}\hat{K}_{M-\{i\}}\otimes\left(\hat{K}_{\{i\}} \hat{S}_i^{(\theta)} + \hat{S}_i^{(\theta)} \hat{K}_{\{i\}}\right).
				\end{align*}
			\end{widetext}
			Taking into account that
			\begin{align*}
				&\bra{S_z^i=z_1}\left(\hat{K}_{\{i\}} \hat{S}_i^{(\theta)} + \hat{S}_i^{(\theta)}\hat{K}_{\{i\}}\right)\ket{S_z^i=z_2} \\
				=&\left[(-1)^{S+z_1}+(-1)^{S+z_2}\right]\bra{S_z^i=z_1}\hat{S}_i^{(\theta)}\ket{S_z^i=z_2} \\
				=&\left[(-1)^{S+z_1}+(-1)^{S+z_2}\right]\delta_{z_1,z_2\pm 1}\bra{S^z_i=z_1}\hat{S}_i^{(\theta)}\ket{S_z^i=z_2} \\
				\equiv&0,
			\end{align*}
			we have
			\begin{align*}
				\hat{O}_i^2 = k^2 \hat{I}_M + \frac{x^2+y^2}{S^4}\left[\hat{S}_i^{(\theta)}\right]^2.
			\end{align*}
			Since the maximum eigenvalue of $\left[\hat{S}_i^{(\theta)}\right]^2$ is $S^2$, we have
			\begin{align*}
				\braket{\hat{O}^2} \le k^2 + \frac{x^2+y^2}{S^2}.
			\end{align*}
			Furthermore, examing the following inequality:
			\begin{align*}
				0 \le& \braket{\hat{O}^2} -\braket{\hat{O}}^2 \\
				\le& \left(k^2+\frac{x^2+y^2}{S^2}\right)-\left(k^2+\frac{x^2+y^2}{S^2}\right)^2 \\
				=& \left(k^2+\frac{x^2+y^2}{S^2}\right)\left[1-\left(k^2+\frac{x^2+y^2}{S^2}\right)\right].
			\end{align*}
			we finally establish:
			\begin{align*}
				k^2+\frac{x^2+y^2}{S^2} \le 1.
			\end{align*}
			and the inequality (\ref{eqn:key}) proved. For the same reason, we deduce that:
			\begin{align*}
				{\left<\hat{K}_{\bar{M}}\right>}^2+\frac{1}{S^2}{\left<\hat{S}^{x}_{j}\right>}^2+\frac{1}{S^2}{\left<\hat{S}_j^{y}\right>}^2 \le 1.
			\end{align*}
			Consequently, for pure bi-separable state $\ket{\phi}$, criterion (\ref{criterion}) is verified. This result can also extends to mixed bi-separable states since the equation $(\hat{K}_M)^{-1} \hat{S}^\pm \hat{K}_M = \hat{S}^\pm$ holds.
		\end{proof}

	\end{document}